\documentclass{article}

\usepackage{PRIMEarxiv}

\usepackage[utf8]{inputenc} 
\usepackage[T1]{fontenc}    
\usepackage[hidelinks]{hyperref}       
\usepackage{url}            
\usepackage{booktabs}       
\usepackage{amsfonts}       
\usepackage{nicefrac}       
\usepackage{microtype}      
\usepackage{amssymb}        
\usepackage{lipsum}
\usepackage{fancyhdr}       
\usepackage{graphicx}       
\usepackage{subcaption}
\usepackage{tabularx}
\usepackage[numbers,sort&compress]{natbib}
\usepackage{amsmath}
\graphicspath{{media/}}     

\title{Stability analysis of thermodiffusively unstable counterflow lean premixed hydrogen flames
}

\author{
  A. Porcarelli$^{a,*}$, P.E. Lapenna$^{b}$, F. Creta$^{b}$ and I. Langella$^{a}$ \\
  $^{a}$Faculty of Aerospace Engineering, TU Delft, Kluyverweg 1, 2629 HS, Delft, Netherlands \\
  $^{b}$Department of Mechanical and Aerospace Engineering, Sapienza, University of Rome, \\ Via Eudossiana 18, 00189, Rome, Italy \\
  \texttt{*\href{mailto:a.porcarelli@tudelft.nl}{a.porcarelli@tudelft.nl}} \\
}

\begin{document}
\maketitle

\begin{abstract}
This study investigates the effect of increasing strain rate on thermodiffusively unstable, lean premixed hydrogen flames in a 2D counterflow configuration through detailed-chemistry numerical simulations for the first time. The analysis of transient flame dynamics without imposed perturbations reveals that a steady-state flame front is achieved only when the strain rate exceeds a certain threshold. Below this threshold, the flame exhibits unstable oscillatory behavior. When subjected to a range of perturbation wavelengths, the flame front exhibits an exponentially increasing wavelength over time, driven by the flame-tangential velocity component, with the applied strain rate acting as the amplification factor. It is shown that any perturbation is damped at sufficiently high applied strain rate conditions after a transient phase. At these high strain regimes, the growth rate transient follows a characteristic onset that depends uniquely on the initial perturbation wavelength and exhibits a linear dependence on the applied strain rate.
\end{abstract}


\section{Introduction}
Hydrogen has emerged as a key contributor in addressing climate change, 
as it offers a zero-carbon alternative to hard-to-electrify sectors like transportation. Recent research efforts have focused on studying hydrogen combustion in lean premixed conditions, where the lower adiabatic flame temperature enables a significant reduction in harmful NO$_{\rm x}$ emissions. In fact, hydrogen's high reactivity 
make it well-suited for achieving ultra-lean regimes without incurring in lean blow off~\cite{cho2009improvement}. However, lean hydrogen flames are also characterised by a very high flame speed, auto-ignition phenomena and growth of thermodiffusive instabilities~\cite{lapenna2023hydrogen}, making flame control and flashback prevention particularly challenging.

Well-established theoretical studies of flame intrinsic instabilities (hydrodynamic and thermodiffusive) are available in literature describing their onset and growth rate at each scale (so-called "dispersion relation") in freely-propagating premixed flame configuration using the hydrodynamic model~\cite{matalon1982flames}, which assumes infinitely thin flame fronts. More simplified theoretical models exist, which under the constant density flow assumption are able to account for lower-than-unity mixture Lewis numbers~\cite{sivashinsky1977diffusional}. Additionally, by incorporating all diffusion effects within a flame speed relation, Creta and Matalon~\cite{creta2011strain} explicitly included strain effects in their model.

Since the recent hit of hydrogen combustion, extensive numerical studies of thermodiffusively unstable flames based on comprehensive modelling have been published~\cite{frouzakis2015numerical,berger2022intrinsic1,al2024efficient}. Along with high-fidelity simulations, recent works attempted to model thermodiffusive instabilities at sub-filter scales of laminar flames with tabulated chemistry to aid the development of low-fidelity models and design practical combustor settings where lean premixed hydrogen flames are stabilised and controlled~\cite{lapenna2024posteriori,remiddi2024data}. Direct numerical simulations~\cite{berger2022synergistic} and experiments~\cite{lapenna2024synergistic} in turbulent conditions further suggested that thermodiffusive instabilities feature a synergestic interaction with turbulence in enhancing hydrogen reactivity and consumption speed, posing further challenges to the accurate modelling of this phenomenon in a large-eddy simulation framework. 

While flame instabilities for unstretched freely-propagating premixed flames have been investigated by numerous numerical and analytical studies, their onset in strained configurations is less understood. Previous works have  shown that hydrogen exhibits a distinctive response to strain, characterised by enhanced overall flame reactivity~\cite{van2016state}, significant delays in the extinction strain rate~\cite{cho2009improvement}, reduction in NO$_{\rm x}$ emissions~\cite{porcarelli2023suppression}, and mitigation of preferential diffusion effects~\cite{porcarelli2024mitigation}. The stability of stagnation-point flames has been only theoretically studied by Sivashinsky et al.~\cite{sivashinsky1982stability},  highlighting that a sufficiently high strain rate may lead to an overall stabilisation of the perturbation modes in all directions. However, stability analyses based on high-fidelity simulations in strained configurations are still to be done, and it has yet to be demonstrated that strain can by itself suppress intrinsic instabilities, thereby aiding the stabilisation of hydrogen flames.

The purpose of this study is to characterise the response of a lean premixed and strained hydrogen flame front to a range of perturbation wavelengths. Numerical simulations with
detailed-chemistry accounting for the relevant transport phenomena in hydrogen flames are performed in 2D laminar counterflow configurations with varying levels of applied strain rate. Results show that above a threshold applied strain rate, after an initial transient where the perturbation grows, the initial and stable unperturbed flame state is always restored.

\section{Computational setup}

\subsection{Governing Equations}
Detailed-chemistry, two-dimensional, laminar reacting flow simulations are performed using an \textit{in-house} version of reactingFoam. This transient compressible solver in OpenFOAM has been modified to incorporate mixture-averaged transport and temperature-dependent thermodynamic and transport properties. The reacting Navier-Stokes equations \cite{poinsot2005theoretical} are solved for mass, momentum, sensible enthalpy $h$ and the mass fraction $Y_k$ of $N-1$ species $k$. In Einstein's notation, the equation for the generic species $k$ is
\begin{equation}
       \label{eq:species}
       \frac{\partial(\rho Y_k)}{\partial t}+\frac{\partial (\rho u_i Y_k)}{\partial x_i} = -\frac{\partial (\rho {V}_{k,i} Y_k)}{\partial x_i} + W_k \dot{w}_k,
\end{equation}

where $\rho$ is the mixture density, $V_{k,i}$ is the diffusion velocity of species k, $\dot{w}_k$ is the molar rate of production of species $k$, $W_k$ is the molar mass of species $k$, and subscripts $i$ and $j$ denote the spacial directions. 
Body forces, viscous dissipation, and pressure gradients are neglected. 
The low-Mach ideal gas law and the caloric equation of state are used as thermodynamic model, where in the latter the species heat capacities are obtained using the JANAF polynomials.
The mixture viscosity is calculated a priori using the kinetic transport data with the TROT code in Chem1D~\cite{chem1d}, following the method of Wilke~\cite{wilke1950viscosity} as described by Evlampiev~\cite{evlampiev2007numerical}. The results are then tabulated as a logarithmic 3rd-order polynomial function of temperature. 

A mixture-averaged diffusion model~\cite{poinsot2005theoretical} with velocity correction is used to model the diffusion velocity and account for lean hydrogen's preferential and differential diffusion effects:
\begin{subequations}\begin{equation}
    \label{eq:fick}
    V_{k,i} = - \frac{D_k^M}{X_k} \frac{\partial X_k}{\partial x_i} + V_{c1,i} \approx \frac{D_k^M}{Y_k} \frac{\partial Y_k}{\partial x_i} + V_{c,i},
\end{equation}
    \begin{equation}
        \label{eq:diffcoeffs}
        D_k^M = \frac{1-Y_k}{\sum_{l,k\neq l}^{N} {X_l}/{D_{kl}}}
    \end{equation}
    \label{eq:diff}
\end{subequations}
where $D_k^M$ are the mixture-averaged diffusion coefficients and $V_{c,i}$ is the correction velocity:
\begin{equation}
    \label{eq:Vc}
    V_{c,i} = \sum_{k=1}^N D_k^M Y_k.
\end{equation} 
The correction velocity follows from the mass conservation to ensure that $\sum_{k=1}^N j_k = 1$, where $j_k$ are the species mass fluxes. The binary diffusion coefficients in Eq.~\eqref{eq:diff} are  computed using the approximation of Hirschfelder et al.~\cite{hirschfelder1964molecular} and tabulated in OpenFOAM as a function of temperature \cite{evlampiev2007numerical}. 
The effect of Soret or thermal diffusion is neglected at this stage and will be investigated in a future study. 

The heat flux is found as the composition of conductive and diffusive contributions
\begin{equation}
    \label{eq:q}
    q_i = q_{i,\rm cond} + q_{i, \rm diff} = -\lambda \frac{\partial t}{\partial x_i} + \rho \sum_{k=1}^{N} h_k Y_k V_{k,i},
\end{equation}
where  mass conservation is ensured by setting $\sum_{k=1}^{N} h_k j_k = 0$, leading to the following expression of the diffusive heat flux
\begin{equation}
    \label{eq:qDiff}
    \begin{split}
        q_{i, \rm diff} = \rho \sum_{k=1}^{N} h_k Y_k V_{k,i} = \\
        \rho \sum_{k=1}^{N-1} h_k Y_k V_{k,i} - \rho h_N \sum_{k=1}^{N-1} Y_k V_{k,i}.
    \end{split}
\end{equation}
Similarly to the mixture viscosity, the mixture conductivity is also determined \textit{a priori} and tabulated as function of temperature. Radiation and Dufour effects are neglected in this study. 
Detailed kinetic data of reactions and species transport are taken from the Conaire chemical mechanism~\cite{conaire2004comprehensive}.

\subsection{Computational setup and numerical details}
\begin{figure}[h!]
\centering
\includegraphics[width=0.48\textwidth]{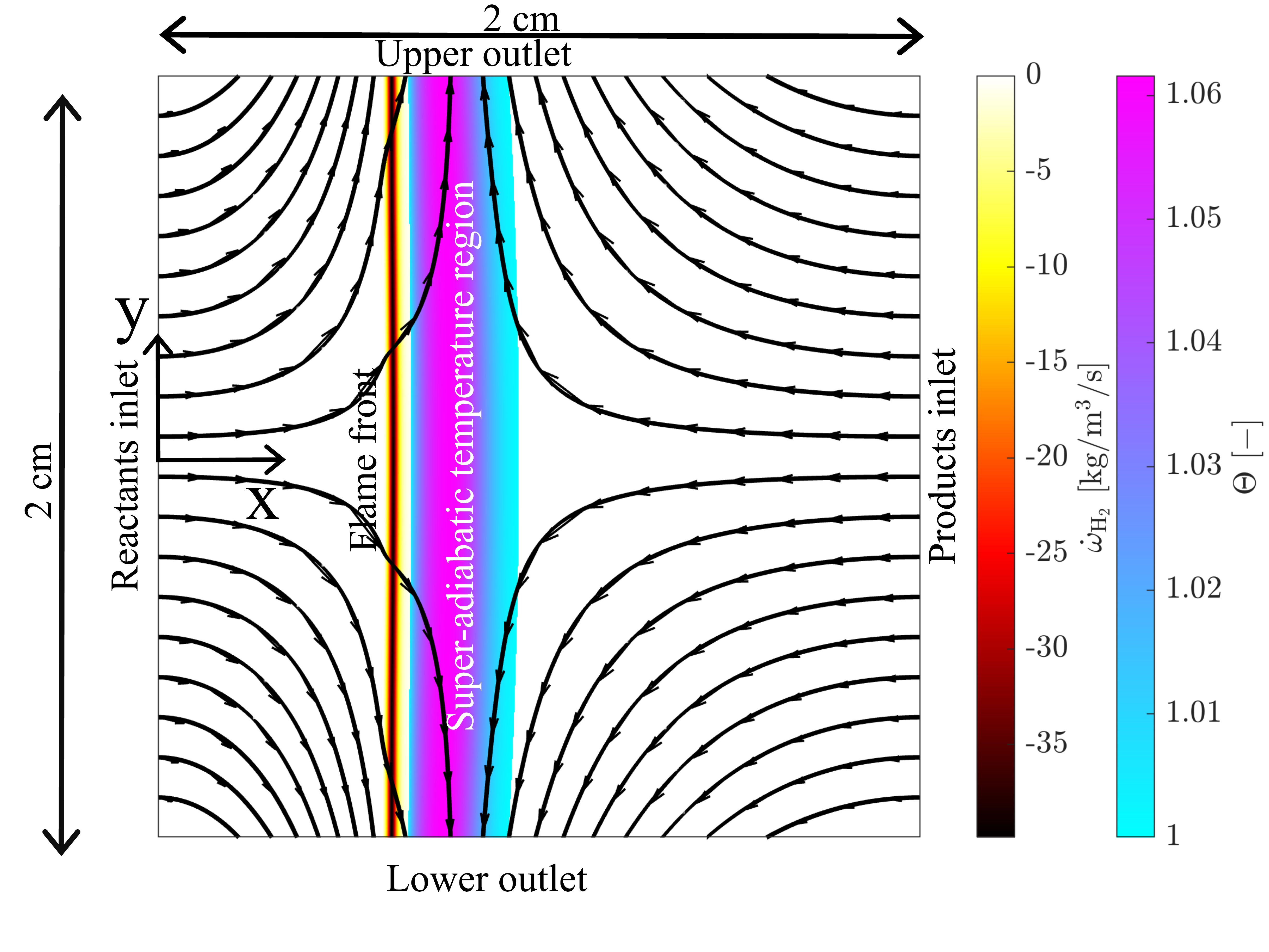}
\caption{\footnotesize Sketch of the two-dimensional counterflow setup.}
\label{fig:CF2D}
\end{figure}

The flame setup consists of a counter-flow reactants-to-products configuration as shown in Fig.~\ref{fig:CF2D}.
As discussed in a recent study~\cite{porcarelli2023suppression}, such configuration is considered the most suitable to analyse lean premixed hydrogen flames at intensive strain rates. Similarly to previous studies performing linear stability analyses of unstretched thermodiffusively unstable flames \cite{al2024efficient,frouzakis2015numerical,berger2022intrinsic1}, lean conditions are established at an equivalence ratio $\phi=0.5$ and the reactants temperature and pressure are fixed respectively to T$_r$=300 K and $p=1$ atm. At the products boundary, the temperature is prescribed to adiabatic conditions, T$_p$=1646 K, and the mixture composition is imposed from complete combustion. 
The nominal applied strain rate in this study is defined as
\begin{equation}
    \label{eq:a}
    a = \frac{\left | u_r \right | + \left | u_p \right |}{L},
\end{equation}
where $L=2$ cm is the domain length, and $u_r$ and $u_p$ are the velocities at the reactants and products boundary, respectively. 
Four cases at different applied strain rate are investigated in the present study and reported in Table~\ref{tab:setup} along with  
their consumption speed $s_c$, thermal flame thickness $\delta_f=\frac{T_p-T_r}{|\partial T / \partial x |_{\rm max}}$, and the chemical time scale $\tau_f=\delta_f/s_c$. 
Different nominal strain rates are achieved by prescribing the velocity at the reactants and products boundary. 
\begin{table}[h!] \footnotesize
\caption{Overview of the four simulations investigated in the present work.}
\centerline{\begin{tabular}{lcccc}
\hline 
Case name &  $a$ [s$^{-1}$] & $s_c$ [m/s] & $\delta_f$ [mm] & $\tau_f$ [ms] \\
\hline
		a500 &  346.3 & 0.596 & 0.402 & 0.675 \\
		a1000 &  706.85 & 0.659 & 0.371 & 0.563 \\
            a2000 &  1447.5 & 0.732 & 0.346 & 0.472 \\
            a5000 &  3633.5 & 0.831 & 0.299 & 0.360 \\
\hline 
\end{tabular}}
\label{tab:setup}
\end{table}

The computations are performed with an implicit second-order Crank-Nicolson discretization scheme for time marching, combined with a third-order cubic scheme for the convective term of all resolved quantities. 
A constant time-step is chosen to ensure a maximum CFL number below $0.2$. The simulations are first run over a long transient (up to $t=50 \tau_f$) from ignition to a basic steady state (if achieved, see the next section). Then, a single-wavelength perturbation is applied to the flame front at steady state and is tracked until the crests exit the domain in the upper outlet, corresponding to a physical time of up to $t=7 \tau_f$. The domain  for all  simulations is discretised using a uniform mesh of 800x800 finite volumes, resulting in a cell spacing $\Delta x = 2.5\cdot 10^{-5}$ m. This mesh enables a resolution of the flame structure comparable to prior studies \cite{frouzakis2015numerical}, with $n_f = \delta_f/\Delta x$ ranging between 12 and 15.

\section{Results and discussion} 
\subsection{Validation} 
Results obtained using the presented methodology in OpenFOAM were preliminarily validated for a one-dimensional freely-propagating premixed hydrogen laminar flame at equivalence ratio  $\phi=0.5$, against those obtained from the well-known code Chem1d~\cite{chem1d}. Results indicated a maximum difference of 2.4\%, 4.5\%, and 5.7\% respectively for the mixture fraction dip across the domain, flame thickness, and consumption speed. For further details on this preliminary validation over the 1D setup, the reader is referred to Section 1 in the supplementary material.

\begin{figure}[h!]
\centering
\includegraphics[width=0.48\textwidth]{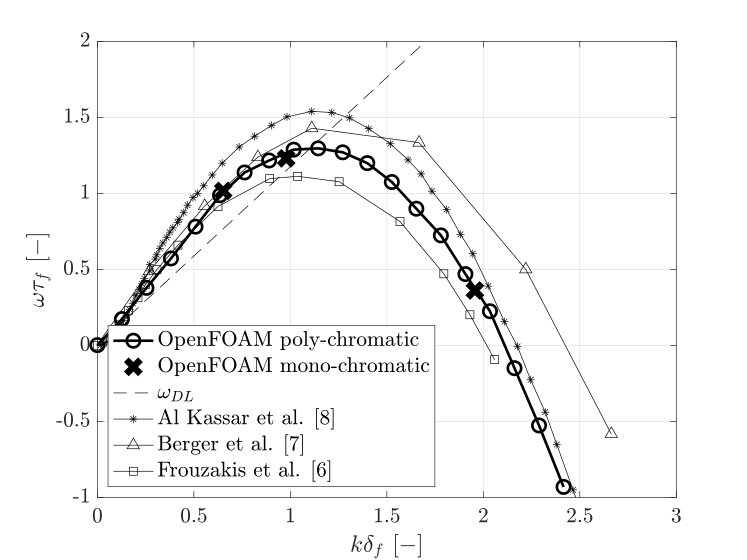}
\caption{\footnotesize Dispersion relation for a freely propagating hydrogen flame obtained from current results, and compared to data from the literature.}
\label{fig:2DValidation}
\end{figure}

In order to assess the solver's capability in performing stability analyses in thermodiffusively unstable flames, a two-dimensional laminar, freely propagating premixed flame with hydrogen fuel at an equivalence ratio of $\phi = 0.5$ is further simulated on a mesh ensuring 25 cells within the laminar flame thickness. The flame is perturbed with a multi-wavelength perturbation with initial amplitude $A_0 = 0.02\delta_f$, and the dispersion relation in the linear regime is reconstructed following the polychromatic methodology in Al Kassar et al.~\cite{al2024efficient}. Results are reported in Fig.~\ref{fig:2DValidation}, and  compared to those in the literature for the same burning regime~\cite{al2024efficient,frouzakis2015numerical,berger2022intrinsic1}.
As observed, the dispersion relation obtained in OpenFOAM  is within the range expected from the other literature studies, indicating that the methodology used in the present work is suitable to study thermodiffusively unstable flames. 

\subsection{Flame dynamics without imposed perturbations} 
\begin{figure}[h!]
\centering
\includegraphics[width=0.48\textwidth]{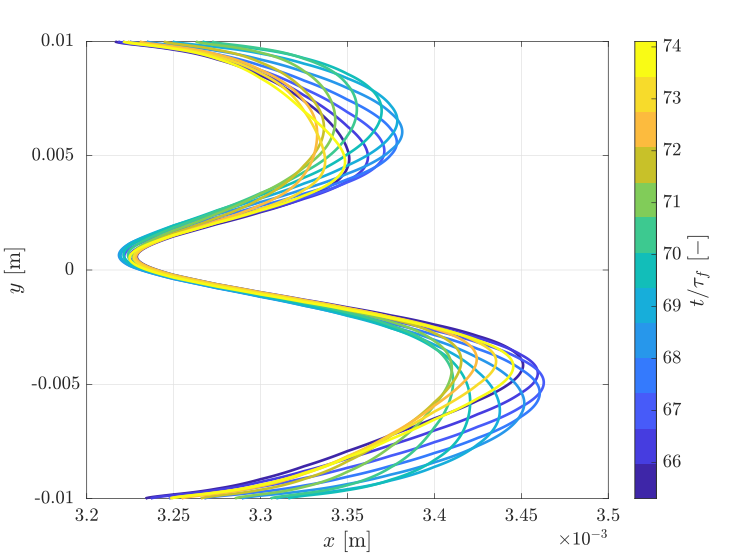}
\caption{\footnotesize  Low strain rate case a500. Evolution of the flame front as a function of time. Note the different x and y axis scale.}
\label{fig:a500FlameFront}
\end{figure}

\begin{figure}[h!]
\centering
\includegraphics[width=0.48\textwidth]{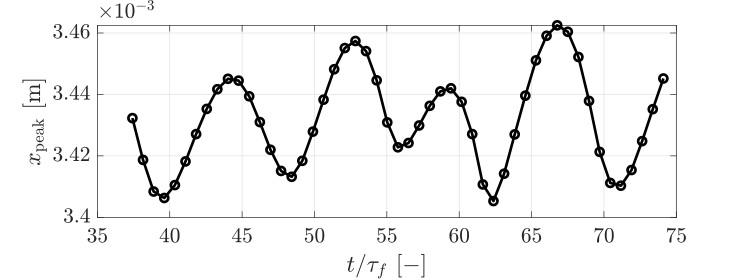}
\caption{\footnotesize Low strain rate case a500. Evolution of the $x$ position of the lower flame tip as a function of time.}
\label{fig:a500FlameTip}
\end{figure}

\begin{figure}[h!]
\centering
\includegraphics[width=0.48\textwidth]{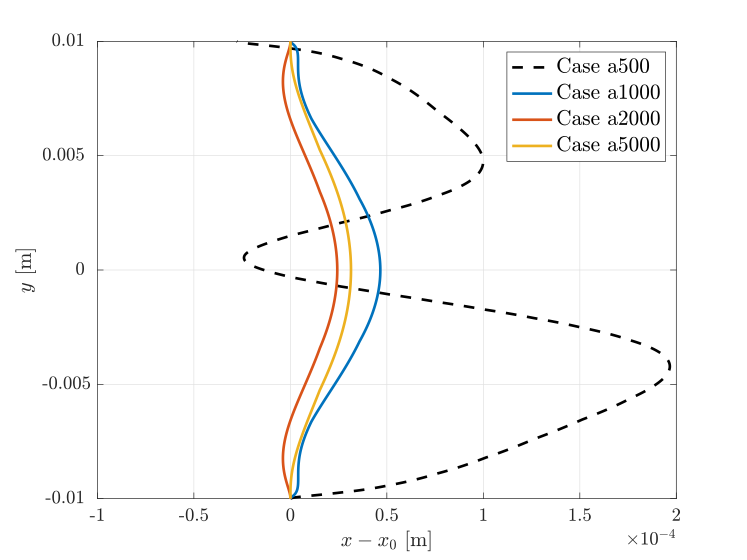}
\caption{\footnotesize Steady flame front shape achieved at medium to high strain rate cases (case a1000 to a5000, solid lines). Comparison to an instantaneous unsteady flame front at low strain rate (case a500, dashed line). Note the different x and y axis scale.}
\label{fig:basicStates}
\end{figure}

The flame dynamics at different applied strain rates without imposing any perturbation yet is discussed first to have a base reference.
At the lowest strain rate, case a500 of Table~\ref{tab:setup}, the flame front does not reach any steady state after an initial transient. As shown in Fig.~\ref{fig:a500FlameFront}, the flame front is curved and features a small-amplitude, undamped oscillatory pattern in time.
This unsteady oscillatory behavior can be further observed in Fig.~\ref{fig:a500FlameTip}, showing the variation in time of the flame position (reported only at the lower boundary for simplicity). 
This oscillatory behaviour is not observed for the other strain rate cases, where after a numerical transient the flame achieves a steady shape in time, which is reported in Fig.~\ref{fig:basicStates}.
Note that, albeit a steady state is reached for those cases, their respective flame fronts exhibit a mild curvature (observe that the axes in the figure are not uniform). This is however due to some variation in the velocity field near the outlet boundaries, rather than being an intrinsic effect of the flame dynamics.
On the contrary, the unsteady a500 case shows an opposite concavity with respect to the other cases. Here, the instantaneous, positively curved flame front $\kappa>0$ (concave towards the products) near the domain centreline is located more upstream than for the higher strain rate cases at any time, implying that the flame is propagating locally at a higher speed. Conversely, the negatively curved flame fronts $\kappa<0$ (concave towards the reactants) at $y\approx \pm 5$ mm are located more downstream at any time step as compared to the higher strain rate cases, indicating that the flame speed is slower at these locations. 
These differences for the a500 case with respect to the cases at higher strain rate are the result of the occurrence of intrinsic instabilities. However, this behaviour is different from that of an unstretched thermodiffusively unstable case, where the flame surface area would keep growing in time and is not expected to exhibit the oscillatory behavior highlighted in Figs.~\ref{fig:a500FlameFront} and~\ref{fig:a500FlameTip}. Indeed, the aforementioned oscillatory behavior is a characteristic feature of the counterflow configuration, where the velocity at the reactants side decreases (about linearly) in the streamwise direction. 
This implies that when a positively (negatively) curved flame front, due to its higher (lower) flame speed, propagates upstream (dowstream) towards the reactants (products), it will also encounter a higher (lower) velocity and lower tangential strain, both counteracting the propagation effect of increased (decreased) flame speed. 
For this case, the intrinsic instability onset and the (stabilizing) effect of the counterflow configuration are of similar magnitude, so that the unsteady oscillatory pattern described earlier is achieved. For the higher strain rate cases the velocity gradient in the streamline direction is stronger, implying that the stabilising effect is dominant over the onset of intrinsic instabilities, thus a steady flame front is achieved.

The above results suggest that for sufficiently high strain levels the counterflow configuration stabilises hydrogen flames that would be thermodiffusively unstable in unstretched conditions. This point will be further discussed in the following sections.

\subsection{Perturbed flame front dynamics} 
\begin{figure}[h!]
\centering
\includegraphics[width=0.48\textwidth]{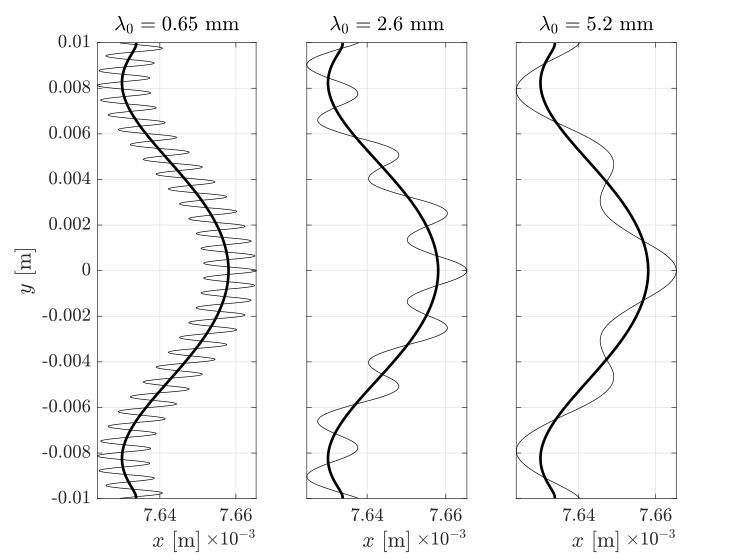}
\caption{\footnotesize Basic state flame front (thick line) vs perturbed flame front (thin line) at different perturbation wave lengths $\lambda_0$ for case a2000 of Table~\ref{tab:setup}.}
\label{fig:perturbation}
\end{figure}

The steady flame front obtained for cases a1000, a2000, and a5000 of Table~\ref{tab:setup} discussed in the previous section was perturbed with a single-wavelength or mono-chromatic signal, and the response of the flame to this perturbation is discussed here. 
The initial range of perturbation wavelength $\lambda_0$ is chosen such that the corresponding wave number $k_0=2\pi/\lambda_0$ falls within the unstable modes in the dispersion relation of unstretched flames for the conditions investigated (see Fig.~\ref{fig:2DValidation}). For completeness, one case ($\lambda_0=0.65$ mm) is also chosen in the negative growth rate dispersion relation region. The applied perturbations are summarised in Table~\ref{tab:perturbation}. The normalised wave number $k_0 \delta_f$ is also reported as a mean of comparison to the dispersion relation for freely-propagating flame in Fig.~\ref{fig:2DValidation}.
\begin{table}[h!] \footnotesize
\caption{Perturbations applied to the basic state flame for cases a1000, a2000 and a5000.  $\lambda_0$ is the initial perturbation wavelength, $k_0$ is the corresponding wave number  and $\delta_{f}$ is the unstretched flame thickness.}
\centerline{\begin{tabular}{ccc}
\hline 
$\lambda_0$ [mm]  & $k_0$ [1/mm] & $k_0\delta_f$ [$-$] \\
\hline
		0.65 & 9.67 & 3.91 \\
		1.3 & 4.83 & 1.96 \\
            2.6 & 2.42 & 0.978 \\
            3.9 & 1.61 & 0.652 \\
            5.2 & 1.21 & 0.489 \\
            6.5 & 0.967 & 0.391 \\
\hline 
\end{tabular}}
\label{tab:perturbation}
\end{table}
The perturbation signal is given to all fields with a cosine function to preserve the flame front symmetry, with an initial amplitude of $A_0 = 0.02 \delta_f$ (similar to previous studies~\cite{berger2022intrinsic1}). An example of how the steady state flame front (basic state) is perturbed in the chosen range of $\lambda_0$ is given in Fig.~\ref{fig:perturbation}.
Once the flame is perturbed, as one would expect the initial (imposed) wavelength increases in time due to the effect of the vertical velocity in the counterflow configuration of Fig.~\ref{fig:CF2D}. To better visualise this phenomenon, the reader is referred to Section 2 of the supplementary material.
\begin{figure*}[h!]
	\centering
        \begin{subfigure}{0.74\textwidth}
                \centering
                \includegraphics[width=\textwidth]{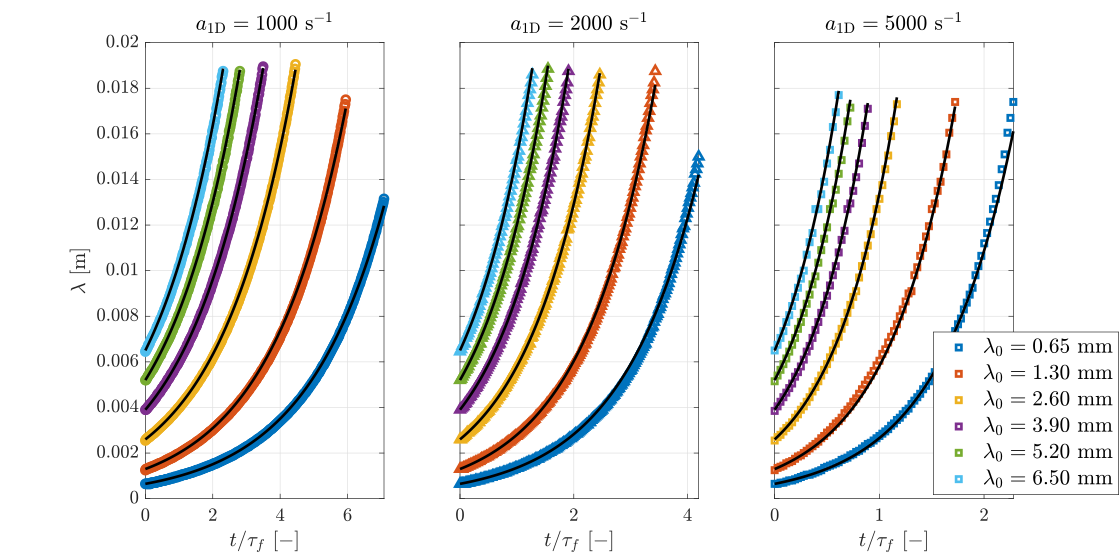}
                \caption{}
                \label{fig:lambdat}
        \end{subfigure}
        \begin{subfigure}{0.245\textwidth}
	       \centering
	       \includegraphics[width=\textwidth]{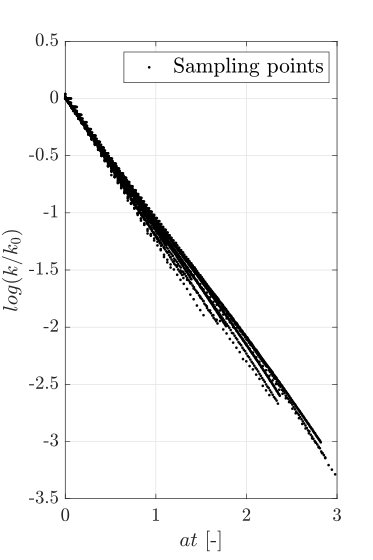}
	       \caption{}
	       \label{fig:logkat}
        \end{subfigure}
	\caption{\footnotesize (a) Comparison between the numerical values of wave length $\lambda$ and the pseudo-analytical expression $\lambda = \lambda_0 e^{K_st}$ for different perturbation wave lengths $\lambda_0$ and strain rates. (b) Collapsing sampling points in $at-{\rm log}(k/k_0)$ space at all perturbation wave lengths $\lambda_0$ and strain rates investigated.}
	\label{fig:lambdatLogkat}
\end{figure*}

Before discussing the evolution in time of $\lambda$, it is however convenient to derive an analytical formulation for its expected shape function. Considering that $A_0<<\delta_{f}$, the local tangential strain rate $K_s$ for the opposed-jet configuration can be reduced to that of a flat flame front,
$K_s = \frac{\partial u_y}{\partial y}$.
%
Let's also assume that $K_s$ is about constant in time along the flame front during its evolution post-perturbation, 
which is confirmed by the numerical results (not shown). 
By replacing $y=\lambda/2$ and recalling that $u_y\simeq\frac{dy}{dt}=y^\prime$ for small perturbations, one then obtains
%
\begin{equation}
    \label{eq:lambdat}
    \lambda (t) = \lambda_0 e^{K_st},
\end{equation}
and in terms of wave number 
\begin{equation}
    \label{eq:kt}
    k(t) = k_0 e^{-K_st}.
\end{equation}
In order to compare this solution with the simulation data, an estimate of the exponential factor $K_s$ is necessary. By noting that $K_s = \frac{d ({\rm log} (\lambda/\lambda_0))}{dt}$ and recalling that $K_s \approx {const}$ in time, the exponential factor is obtained by linear interpolation of the function ${\rm log}\, \frac{\lambda(t)}{\lambda_0}$ obtained from the simulations. 
A comparison between the analytical Eq.~\eqref{eq:lambdat} with fitted $K_s$ and the simulation data for $\lambda(t)$ is shown in Fig.~\ref{fig:lambdat}.
One can observe a very close agreement between the pseudo-analytical expression and the simulation data,  implying the assumptions used for the analytical derivation hold.
It is worth to note that $K_s$ is directly proportional to the applied strain rate $a$. In fact
\begin{equation}
    \label{eq:log(kk0)}
    {\rm log} k(t) = {\rm log}k_0 -K_s t \Rightarrow {\rm log} (k/k_0) = -K_st,
\end{equation}
and thus ${\rm log} \frac{k}{k_0} (K_st)$ is a straight line with slope -1 for all the investigated cases. Should $K_s$ be directly proportional to $a$, the same straight line would be obtained for the ${\rm log} \frac{k}{k_0} (at)$. This is confirmed with very good approximation by the plot in Fig.~\ref{fig:logkat}. This aspect is important as it demonstrates that the mono-chromatically perturbed flame front in a counterflow configuration yields an exponentially increasing wavelength in time with applied strain rate as amplification factor. Some deviations from the pseudo-analytical expression are found because the local $K_s$ shows moderate variations depending on the local perturbations, which are locally affecting the flow field on the small scales. Yet, the fit remains linear with very good approximation if the global applied strain rate is considered.

\subsection{Perturbation growth rate} 
\begin{figure*}[h!]
\centering
\vspace{-0.4 in}
\includegraphics[width=\textwidth]{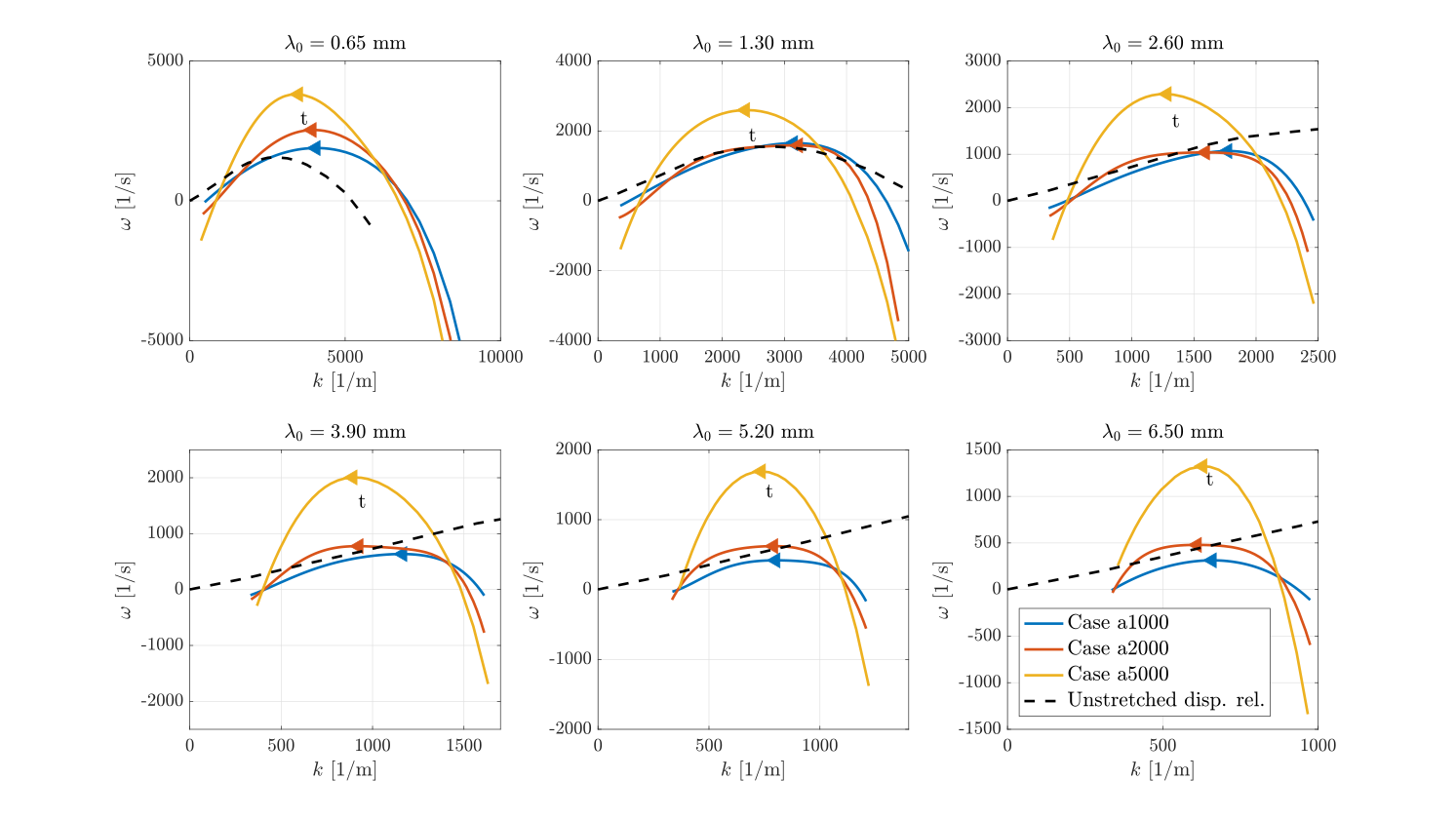}
\vspace{10 pt}
\caption{\footnotesize Growth rate $\omega$ as a function of the wave number $k$ at different strain rates and different initial perturbation wave lengths $\lambda_0$.}
\label{fig:omegakBis}
\end{figure*}

The behavior of the perturbation growth rate $\omega$ is investigated in this section for the three stable cases of Table~\ref{tab:setup} (a1000 to a5000). The growth rate is defined as:
\begin{equation}
    \label{eq:omega}
    \omega = \frac{d ({\rm log} A)}{dt} = \frac{1}{A} \frac{dA}{dt},
\end{equation}
where $A$ is the time-evolving amplitude. Growth rates are shown for different initial wavelengths and applied strain rates in Fig.~\ref{fig:omegakBis} as a function of the wave number. The dispersion relation for the unstretched case is also reported for comparison purposes (dashed lines). Note that in order to suppress some noise from the simulation data, the counterflow curves in the figure have been smoothened out using a fifth-order polynomial fit over the sampling points, which are the same as shown in Fig.\ref{fig:lambdat} (one every 0.01 ms). The lowest value of $k$ on the $x$-axis in the plots represents the computed wavelength at the time the last remaining crest reaches the upper outlet of the domain.
It is worth mentioning that these plots do not display a dispersion relation. An unstretched premixed flame under a similar monochromatic perturbation exhibits a constant wave number over time in the initial linear regime, corresponding to a constant growth rate and thus enabling the construction of a dispersion relation. However, in strained configurations, $k$ decreases in time according to Eq.~\eqref{eq:kt}. Consequently, $\omega$ is also not constant for a given $\lambda_0$, as no initial linear regime can be identified due to this mode shift. Note that this is also why, unlike for the commonly used dispersion relation that can be represented with good accuracy with a fourth-order polynomial~\cite{berger2022intrinsic1}, fitting $\omega(k)$  for the strained configurations required a fifth-order polynomial.

For all cases, similar growth rate patterns can be recognised: $\omega$ initially exhibits negative values for high values of $k$ (recall that in time the graph should be read from right to left) and increases with (decreasing) $k$ until it becomes positive and reaches a maximum for a certain value $k^*$. For lower values of $k$, i.e. further progressing in time, $\omega$ is then observed to decrease until it becomes negative again. 
This result indicates that, after an initial and temporary growth, any perturbation is always damped in the conterflow configuration as long as sufficient time has passed. In other words, sufficiently high applied strain rates (let's recall that case a500 of Table~\ref{tab:setup} does not reach a steady state) suppress the intrinsic instability onset, regardless of the initial perturbation wavelength. Note that the temporary positive growth rate region is relatively short, as it spans over a time up to $7\tau_f$.

\begin{figure}[h!]
	\centering
	\includegraphics[width=0.48\textwidth]{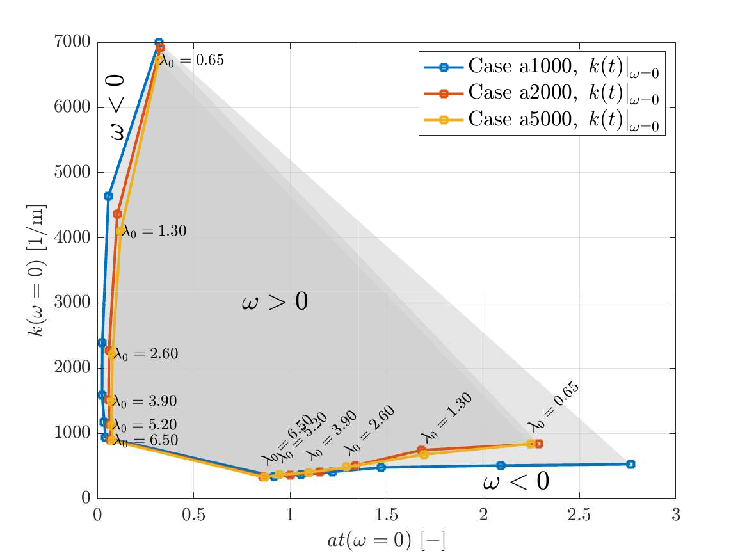}
	\caption{\footnotesize Isolines of growth rate $\omega=0$ in the $at-k$ space for different strain rates and initial perturbation wave lengths $\lambda_0$ [mm]. The set of points on the left-hand-side of the figure represent the `earlier' roots of $\omega(k)$ (refer to the roots on the right of $\omega(k)$ in Fig.~\ref{fig:omegakBis}), while the set of points on the bottom-right side represent the `later' roots (refer to the roots on the left of $\omega(k)$ in Fig.~\ref{fig:omegakBis}).}
	\label{fig:k0at0}
\end{figure}

Further insight on the role of strain rate in suppressing the intrinsic instabilities onset is provided next. By combining Eqs.~\eqref{eq:omega} and~\eqref{eq:kt}, the growth rate can be rewritten as
\begin{equation}
    \label{eq:omegaDecomp}
    \omega = \frac{1}{A} \frac{dA}{dk} \frac{dk}{dt} = - \frac{1}{A} \frac{dA}{dk} K_s k.
\end{equation}
The above expression indicates that the growth rate in $k$ space would vary linearly with the strain $K_s$ if the derivative $\frac{dA}{dk}$ is independent of strain. To assess whether this is the case, one can compute the roots of $k$ at zero growth rate, $k(\omega=0)$. By looking at Eq.~\eqref{eq:omegaDecomp}, since for the cases investigated $K_s$ and $k$ are always greater than zero, $\omega=0$ only for $\frac{dA}{dk}=0$. The values of $k$ at zero growth rate, $k(\omega=0)$, are plotted for the corresponding time $t(\omega=0)$ in Fig.~\ref{fig:k0at0}. At each initial perturbation $\lambda_0$ there corresponds a pair of $k(\omega=0)$ points on the graph. 
Furthermore, two distinct sets of points can be identified. On the left-hand side, the `earlier' roots of $\omega(k)$ are shown, which correspond to the points in Fig.\ref{fig:omegakBis} where the growth rate transitions in time from negative to positive. In contrast, the `later' roots are located at the bottom-right, corresponding to the points in Fig.\ref{fig:omegakBis} where the growth rate shifts from positive to negative, ultimately suppressing the instabilities over time.
The region enclosed by the iso-line (shaded in Fig.~\ref{fig:k0at0}) represents the range of modes and times for which the growth rate is positive.
\begin{figure*}[h!]
\centering
\vspace{-0.4 in}
\includegraphics[width=\textwidth]{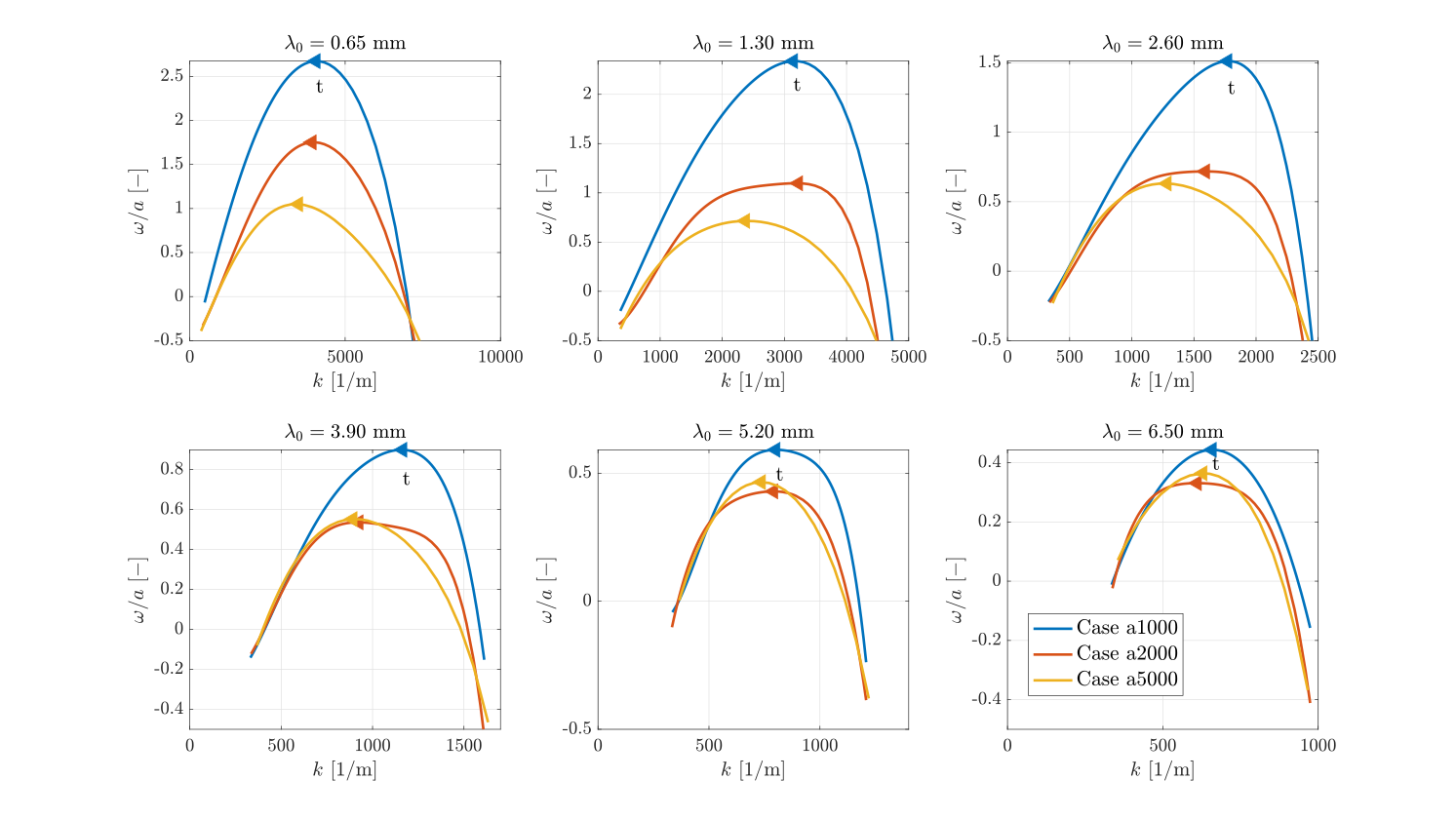}
\vspace{10 pt}
\caption{\footnotesize Growth rate $\omega$ as a function of the wave number $k$ normalised by the applied strain rate $a$ at different initial perturbation wave lengths $\lambda_0$.}
\label{fig:omegakTer}
\end{figure*}
\\
This figure highlights that the region of instability at the two higher strain rates, cases a2000 and a5000 of Table~\ref{tab:setup}, almost perfectly overlap, while the same region of instability is broader at the lower strain rate case. In particular, the `later' roots are closer to $k=0$ in the a1000 case. This suggests that a further decrease of applied strain would virtually shift the values of $k(\omega=0)$ at $k<0$, meaning that there would not be any stabilisation over time. This is consistent with the discussion in the previous section, where an unsteady evolution of the flame front (which can be considered equivalent to a random perturbation) was observed for the relatively low strain case (case a500 of Table~\ref{tab:setup}). In that case, in fact, strain is not sufficiently high and a succession of positive and negative growth rates occurs as time progresses, as shown by Fig.~\ref{fig:a500FlameTip} where the amplitude of the local bumps in the flame front is observed to alternatively grow (due to intrinsic instabilities) and shrink (due to the stabilising effect of the flow field in the counterflow). As suggested by means of analytical analyses by Sivashinsky et al.~\cite{sivashinsky1982stability} for stagnation point flames, a further decrease in strain levels is expected to result in growing instabilities similar to the unstretched premixed flame case. However, this instability would exhibit a non-linear onset due to the influence, albeit not dominant, of the tangential velocity component. Note that, although Sivashinsky et al.~\cite{sivashinsky1982stability} analyse a different configuration, similar behaviours to those in the reactants-to-products setup can be expected at low strain, as the flame remains sufficiently distant from the wall or stagnation plane. However, at higher strain rates, stagnation point flame analyses may not be reliable due to unpredictable flame-wall interactions. For sufficiently high strain rates, instead, the observed growth rate pattern characterised by the shift of modes triggered by the tangential velocity becomes increasingly dominant over the single-mode constant growth rates typical of unstrethced conditions, resulting in the stabilisation of the flame.

It is also worth noting that the collapse of the $\omega=0$ iso-lines in the $k-at$ space for the two higher strain rate cases (a2000 and a5000) implies that the equation $\frac{dA}{dk}=0$ is independent of strain in the limit of high applied strain rates. This suggests that $\omega(k)$ may become a linear function of $K_s$ or $a$ (recall Eq.~\eqref{eq:omegaDecomp}) in this limit. This consideration can be verified by looking at the growth rates normalised by the applied strain rate in Fig.~\ref{fig:omegakTer}.
The figure shows that for the two higher strain rates, the curves overlap for the majority of the $\lambda_0$ cases, whereas no such overlap is observed for the lower strain rate case a1000. 
This observation confirms that at sufficiently high strain rate, the growth rate in $k$ space becomes a linear function of strain rate itself, with the derivative $\frac{dA}{dk}$ becoming independent on $K_s$ or $a$. Hence, for a given initial perturbation $\lambda_0$, a characteristic $\omega(k)$ proportional to the applied strain rate can be identified in the counterflow configuration. 

To understand the physics behind this phoenomenon, it is convenient to define two distinct time scales. The rate of change of the wavelength due to the counterflow characteristic vertical velocity component, $\frac{d\lambda}{dt}$, defines a mode-shifting time scale, $\tau_{\rm MS}$. Meanwhile, $\tau_{\rm DR}$ represents the time required for a given mode to adapt its growth rate to its characteristic unstretched dispersion relation $\omega(k)$. At high strain levels $\frac{d\lambda}{dt}$ is very high and determines a $\tau_{\rm MS}$ that is much shorter than $\tau_{\rm DR}$. This implies that before a perturbation can be affected by the characteristic $\omega(k)$ at unstretched conditions, it has already transitioned to another mode, and $\omega(k)$ will thus follow a characteristic pattern uniquely defined by the counterflow configuration at the given $\lambda_0$ and applied strain rate. In the a1000 case, instead, $\frac{d\lambda}{dt}$ is still low enough to allow the unstretched characteristic $\omega(k)$ to partially sum up to the characteristic counterflow growth rate and ultimately affect the total $\omega(k)$, triggering overall higher transient growth rates in the region of instability. Note that, for the two lowest $\lambda_0$ cases, one can observe a non-perfect scaling also for the a2000 case when $k\gtrsim1500$ m$^{-1}$. This can be explained by the fact that, at earlier times, smaller initial perturbations exhibit lower values of $\frac{d\lambda}{dt}$ compared to wider initial perturbations (see Fig.~\ref{fig:lambdat}), resulting in a larger $\tau_{\rm MS}$ even at higher strain rates.

Overall, the analysis carried in the present study shows that from moderate strain rate upwards, intrinsic instabilities are always damped after sufficient time in a counterflow configuration. At further increased strain rates, the transient shape of $\omega(k)$ becomes completely unaffected by the characteristic growth rate of each mode in unstretched conditions, and is uniquely determined by the applied strain rate and the initial imposed perturbation wavelength.

\section{Conclusions}
Detailed chemistry, two-dimensional simulations have been conducted on pure hydrogen lean premixed flames in counterflow configuration at different strain rates. The response of the flame to ranging wavelengths perturbations has been assessed and the following regimes can be identified considering the onset of intrinsic instabilities:
\begin{itemize}
    \item At very low strain rates, the mode shift time scale is greater than the dispersion relation time scale ($\tau_{\rm MS}>\tau_{\rm DR}$). As suggested analytically by Sivashinsky et al.~\cite{sivashinsky1982stability} for stagnation point flames, intrinsic instabilities always grow similarly to an unstretched premixed flame case at the same conditions, because the perturbation has enough time at a given mode to be strongly influenced by the destabilising growth rate of the dispersion relation $\omega_{\rm DR}(k)$. However, $\omega$ would still follow a non-linear onset due to the counterflow-characteristic presence, albeit not dominant, of a flame-tangential velocity component triggering a slow mode shifting.
    \item At low to moderate strain rates, the mode shift time scale is comparable to the dispersion relation time scale $\tau_{\rm MS} \approx \tau_{\rm DR}$. Here, the intrinsic instabilities onset triggered by the destabilising growth rates from the unstretched dispersion relation and the stabilising effect of the mode shifting in the counterflow define an unstable equilibrium, where a repetitive pattern is established featuring first a growth and then a damping of the perturbation.
    \item At moderate to medium strain rates, the mode shift time scale is smaller than the dispersion relation time scale $\tau_{\rm MS} < \tau_{\rm DR}$. From this level of strain onward, any forced perturbation is damped after sufficient time, as the perturbation switches from a mode to another before the mode has time to adapt its growth rate to the one of the unstretched dispersion relation. Nevertheless, $\omega_{\rm DR}$ maintains a mild destabilising influence on the growth rate in $k$ space, such that $\omega(k)$ is not yet a linear function of strain. 
    \item at high to very high strain rates, the mode shift time scale is much smaller than the dispersion relation time scale $\tau_{\rm MS} << \tau_{\rm DR}$. Here, not only any perturbation is always damped after sufficient time, but also the transitory growth rate evolution is independent on $\omega_{\rm DR}$, establishing a pattern that is uniquely defined by the perturbation wavelength and linearly dependent on the applied strain rate.
\end{itemize}
Overall, this study demonstrates for the first time using high-fidelity numerical simulations that intrinsic instabilities in a counterflow configuration are suppressed by sufficiently high applied strain rates, following a transient onset uniquely defined by both the initial perturbation and the applied strain itself. Future work will aim to extend this analysis to three-dimensional laminar and turbulent counterflow cases, laying the groundwork for stabilizing lean premixed hydrogen flames under strained conditions in practical combustion systems.

\section*{Acknowledgments}
AP and IL acknowledge the Dutch Ministry of Education and Science for providing project funding support via the Sector Plan scheme. AP further acknowledges financial support to perform Short Term Scientific Mission (STSM) at Sapienza (Rome) from the CYPHER consortium 
(COST ACTION CA22151). IL further gratefully acknowledges financial support from the ERC Starting Grant OTHERWISE, grant n. 101078821. PEL and FC acknowledge financial support by ICSC (Centro Nazionale di Ricerca in HPC, Big Data and Quantum Computing) funded by the European Union – NextGenerationEU. The authors also acknowledge the use of the Dutch National Supercomputer Snellius 
grant EINF-10039 to perform the simulations.

\bibliographystyle{unsrtnat}
\bibliography{references}  

\end{document}